\documentclass[11pt,a4paper]{article}
\pdfoutput=1
\usepackage{jheppub}
\usepackage{lineno}
\usepackage{graphicx} 
\usepackage{xcolor}
\usepackage{amsmath}
\usepackage{amssymb}
\usepackage{hyperref}
\hypersetup{
    colorlinks=true,
    linkcolor=blue,
    filecolor=magenta,      
    urlcolor=cyan,
    pdftitle={Overleaf Example},
    pdfpagemode=FullScreen,
    }
\usepackage{comment}
\usepackage{bbold}
\usepackage{tikz}
\usetikzlibrary{arrows}
\usepackage{boxhandler}
\usepackage{subcaption}
\usepackage{multirow}
\usepackage{longtable}

\newcommand{\tr}[1]{\text{Tr}\left(#1\right)}
\newcommand{\llog}[1]{\log{\left(#1\right)}}

\title{Numerical Verification of Perturbative Schwinger-Dyson Resummation on Lattice Models}
\author[a]{Thomas Banks}
\author[b]{Anish Suresh}

\affiliation[a]{New High Energy Theory Center, Rutgers University - New Brunswick, 136 Frelinghuysen Rd, Piscataway, United States}
\affiliation[b]{Rutgers University - New Brunswick, 136 Frelinghuysen Rd, Piscataway, United States}

\emailAdd{tibanks@ucsc.edu}
\emailAdd{anish.suresh@rutgers.edu}

\abstract{We investigate an approximation to the Schwinger-Dyson (SD) equations of the collective Coulomb field of the large $N$ Homogeneous Electron Fluid. The large $N$ approximation transforms the infinite SD hierarchy into a set of closed, equations for 1 and 2-pt correlators. In this paper, the dynamics of a toy model --- a small, square Euclidean lattice with periodic boundary conditions --- are considered. The Markov Chain Monte Carlo numerical method evaluated the 1 and 2-pt correlation functions on a $2 \times 2$ and $3 \times 3$ lattice. The derived equations are checked with the correlator values, and an agreement at $N \sim 100$ to order $10^{-3}$ was found. The agreement can be further strengthened by increasing runs in the Markov Chain Monte Carlo method.}


\arxivnumber{2509.08501}

\begin{document}

\maketitle

\flushbottom

\section{Introduction}

In \cite{tb1}, one of us presented a method for re-summing perturbative and vector $1/N$ expansions by using Schwinger-Dyson (SD) equations. The basic problem with SD equations is that they form an infinite hierarchy. The observation of \cite{tb1} was that, when written in terms of one particle irreducible (1PI) correlation functions, in a model with a small coupling, this hierarchy could be systematically truncated by noting that the $k$ point 1PI correlator vanishes like $g^k$. Each level of truncation gives rise to a finite closed system of non-linear integral equations, which are amenable to solution on modern computers. The same methods apply to the SD equations for the collective fields of vector $1/N$ expansions, where they are a generalization and improvement of a technique proposed in \cite{banks2020systematic}.  

The purpose of the present note is to explore the accuracy of the proposed approximation on simple finite integrals, where we can perform both an ``exact" evaluation of relevant correlators, and solve the approximate equations. We choose a model that is a cartoon of the Homogeneous Electron Fluid, in which the electrons propagate on a $ 2 \times 2$ or $3 \times 3$ lattice. This model is treated in the re-summed $1/N$ expansion, where we choose the truncation of the SD equations in which only the single fermion loop contribution to the three and four point functions of the collective Coulomb field are kept. The unknown correlators that are solved for are the one and two point functions.   

The paper is structured as follows. In section \ref{sec: derivation}, we derive the truncation of the infinite hierarchical equations up to order $1/N$. Then, we describe the numerical method needed to evaluate the correlators in section \ref{sec: numerical}. We also thoroughly explain the specifics of the numerical technique when applied to our problem. Finally, the results of the code, along with the code, are presented in section \ref{sec: results}. We find agreement at the $10^{-3}$ accuracy level for $N \sim 100$ on the $ 2 \times 2$ lattice and on the $3 \times 3$ lattice. While this accuracy is already strong, we also explain how to increase it.

\section{Derivation of Closed Equations}
\label{sec: derivation}

\begin{figure}[t]
    \centering
    \begin{tikzpicture}

        \node (A) at (0,0) {$z_{11}$};
        \node (B) at (0,-3) {$z_{21}$};
        \node (C) at (0,-6) {$z_{11}$};
        \node (D) at (3,0) {$z_{12}$};
        \node (E) at (3,-3) {$z_{22}$};
        \node (F) at (3,-6) {$z_{12}$};
        \node (G) at (6,0) {$z_{11}$};
        \node (H) at (6,-3) {$z_{12}$};
        \node (I) at (6,-6) {$z_{11}$};

        \draw (A) -- (B);
        \draw[dashed] (C) -- (B);
        \draw (D) -- (E);
        \draw (A) -- (D);
        \draw[dashed] (D) -- (G);
        \draw (E) -- (B);
        \draw[dashed] (F) -- (E);
        \draw[dashed] (H) -- (E);
        \draw[dashed] (C) -- (F);
        \draw[dashed] (F) -- (I);
        \draw[dashed] (H) -- (I);
        \draw[dashed] (G) -- (H);
    \end{tikzpicture}
    \caption{A $2 \times 2$ grid with periodic boundary conditions. The sites only connected via dashed lines represent mirrored values, so there are only $4$ unique sites to consider. When generalized, this means that a $\sqrt{M} \times \sqrt{M}$ grid will only possess $M$ unique sites to consider.}
    \label{tikz: grid}
\end{figure}
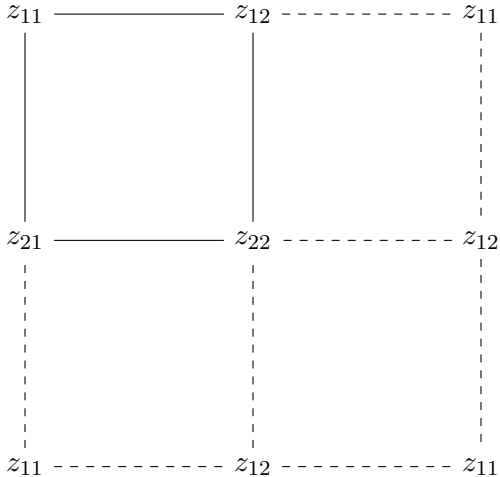

The value of the singlet boson field at each of the $M$ independent lattice sites, $\phi_i$, is simply the stationary point $\Phi_0$ plus a quantum fluctuation term. That is, $\phi_i = \Phi_0 + x(i)/\sqrt{N}.$ These sites are located on a $\sqrt{M} \times \sqrt{M}$ Euclidean lattice with periodic boundary conditions (see figure \ref{tikz: grid}), making the system translational invariant. This also allows us to find a solution for $\Phi_0$ that is independent of the lattice site. It is evident from the definition of $\phi_i$ that the classical regime is recovered when $N \rightarrow \infty$. The action of this model is the following:
\begin{equation}
    \label{eqn: action}
    S = N\bigg[\frac{1}{2} \sum_{i = 1}^M \left(\Phi_0(i) + \frac{x(i)}{\sqrt{N}}\right)^2 + \tr{\llog{A_{ij} + \frac{x(i)}{\sqrt{N}} \mathbb{1}_{ij}}}\bigg].
\end{equation}
Throughout this manuscript, $A_{ij} \equiv K_{ij} + \Phi_0 \mathbb{1}_{ij}$, $K_{ij}$ is the $M \times M$ 2D discrete Laplacian matrix with periodic boundary conditions, and $(A^{-1})_{ij}$ is the propagator between lattice sites $i$ and $j$. With appropriate choices for the matrix $K_{ij}$ this model could approximate a large number of continuum bosonic field theories, with either relativistic or non-relativistic kinematics for the bosons and repulsive point-like interactions. The determinant comes from integrating out $N$ identical bosonic fields.  If we changed the sign of the logarithm it would instead describe fermionic models like the large $N$ Hubbard model or the lattice version of the Gross-Neveu model.  We could also replace the ultra-local quadratic term of the $x(i)$ variables with a lattice Laplacian or Helmholtz operator, to describe particles interacting via Coulomb or Yukawa forces.  In this paper, since we are just testing the numerical accuracy of our proposed approximation method, we will restrict attention to bosons and choose $K_{ij}$ to be a lattice approximation to a two dimensional Laplacian.  Thus, we will be studying the large $N$ relativistic $(\vec{\phi}^2)^2$ model on a very small two dimensional lattice.

\begin{figure}[t]
\centering
    \begin{tikzpicture}[scale=1.2]

    \draw (0,0) ellipse (10mm and 5mm);

    \draw (0,0) node {$S^{(2)}$};

    \draw[dashed] (-1.7,0) -- (-1,0);
    \draw[dashed] (1,0) -- (1.7,0);

    \draw (4,0) node {$S^{(3)}$};

    \draw (4-0.75,-0.433) -- (4+0.75, -0.433) -- (4, 0.866) -- cycle;

    \draw[dashed] (4-0.75,-0.433) -- (4-0.75-0.49,-0.433-0.49);
    \draw[dashed] (4+0.75,-0.433) -- (4+0.75+0.49,-0.433-0.49);
    \draw[dashed] (4, 0.866) -- (4, 0.866+0.7);

    \draw (8-0.5, -0.5) -- (8+0.5, -0.5) -- (8+0.5, 0.5) -- (8-0.5, 0.5) -- cycle;
    \draw (8,0) node {$S^{(4)}$};

    \draw[dashed] (8-0.5, -0.5) -- (8-0.5-0.7, -0.5-0.7);
    \draw[dashed] (8+0.5, -0.5) -- (8+0.5+0.7, -0.5-0.7);
    \draw[dashed] (8-0.5, 0.5) -- (8-0.5-0.7, 0.5+0.7);
    \draw[dashed] (8+0.5, 0.5) -- (8+0.5+0.7, 0.5+0.7);
    
\end{tikzpicture}
\caption{Feynman diagrams for $S^{(2)}$, $S^{(3)}$, and $S^{(4)}$. The propagator here is $A^{-1}_{ij}$.}
\label{tikz: feynman}
\end{figure}
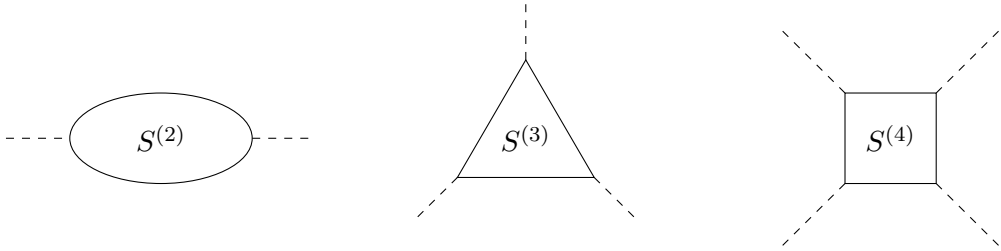

We can then expand the action perturbatively, since we are interested in the terms that dominate in the action for large $N$. Ignoring the constant and linear terms, the action takes the form:
\begin{equation}
    S = \sum_{k = 2}^\infty \frac{N^{1 - k/2}}{k!} \sum_{i_1, \cdots, i_k}^M S^{(k)}(i_1, \cdots, i_k) x(i_1) \cdots x(i_k),
\end{equation}
where $S^{(2)}(i,j) = \delta_{ij} - (A^{-1})_{ij}(A^{-1})_{ji}$ and $S^{(k)}(i_1,\cdots,i_k) = (-1)^{k-1}(k-1)! (A^{-1})_{i_1 i_2}$ $(A^{-1})_{i_2 i_3}$ $\cdots (A^{-1})_{i_k i_1}$. The perturbative expansion provides a straightforward method to determine the stationary point, which we will cover in detail in section \ref{sec: results}. Figure \ref{tikz: feynman} contains the Feynman diagrams for $S^{(2)}, S^{(3)},$ and $S^{(4)}$. Diagrams for higher order terms simply follow this pattern.

We can then write the partition function in the imaginary time formalism/after Wick rotation:
\begin{equation}
    Z = \frac{1}{Z_0} \int [dx] \exp{\left(-S - \sum_i^M x(i) j(i)\right)} = e^{-W}.
\end{equation}
The chain rule with any fixed $i$ states that $\partial_{x(i)} Z = 0$. Only keeping terms up to order $1/N$ and rewriting the partition functions with respect to $W$ and its derivatives, we arrive at the following $M$ equations (one for each $i$):
\begin{equation}
    \begin{split}
        \label{eqn: master1}
        -j(i) = & \sum_{k}^M S^{(2)}(i,k) W_k + \sum_{k,l}^M \frac{1}{2\sqrt{N}} S^{(3)}(i,k,l) (W_k W_l - W_{kl}) \\ 
        & - \sum_{k, l, m}^M \frac{1}{6N} S^{(4)}(i,k,l,m) \bigg[W_{kl}W_m + W_{km}W_l + W_{lm}W_k - W_{klm} - W_{k}W_lW_m\bigg]
    \end{split}
\end{equation}
where subscripts $i_1,\cdots,i_k$ on $W$ indicate derivatives with respect to $j(i_1) \cdots j(i_k)$. Taking the derivative with respect to $j(t)$ on both sides produces another $M^2$ equations (one for each $(i,t)$ pair):
\begin{equation}
    \label{eqn: master 2}
    \begin{split}
        0 = & \sum_{k}^M S^{(2)}(i,k) W_{kt} + \sum_{k,l}^M\frac{1}{2\sqrt{N}}S^{(3)}(i,k,l)\bigg[W_{kt} W_l + W_k W_{lt} -W_{klt}\bigg] + \delta_{it}  \\
        & - \sum_{k,l,m}^M\frac{1}{6N}S^{(4)}(i,k,l,m)\bigg[(W_{klt}W_m + W_{kl}W_{mt} + \text{symmetric terms}) - W_{klmt} \\
        & \hspace{3.65cm} - (W_{kt}W_l W_m + \text{symmetric terms}) \bigg].
    \end{split}
\end{equation}
Only $M(M+1)/2$ out of the $M^2$ equations are unique, since derivatives commute. eqs. \eqref{eqn: master1} and \eqref{eqn: master 2}, which consist of  $M + M(M+1)/2$ equations in total, serve as the metrics for how successful this approximation is. Note that the sources $j(i)$ are set to 0 in eq. \eqref{eqn: master1} by definition.

\subsection{Approximating Higher Order Correlation Functions}

By definition, we have that $W_i$ is the 1-pt correlation function $\langle x(i) \rangle$. There are only $M$ such terms, so these terms are manageable. Similarly, finding the 2-pt correlation function $\langle x(i) x(j) \rangle$ will help us find the connected correlation function $W_{ij}$:
\begin{equation}
    W_{ij} = W_i W_j - \langle x(i) x(j) \rangle.
\end{equation}
However, working with 3 and 4-pt correlation functions is a lot more computationally intensive, especially given the large number of degrees of freedom.

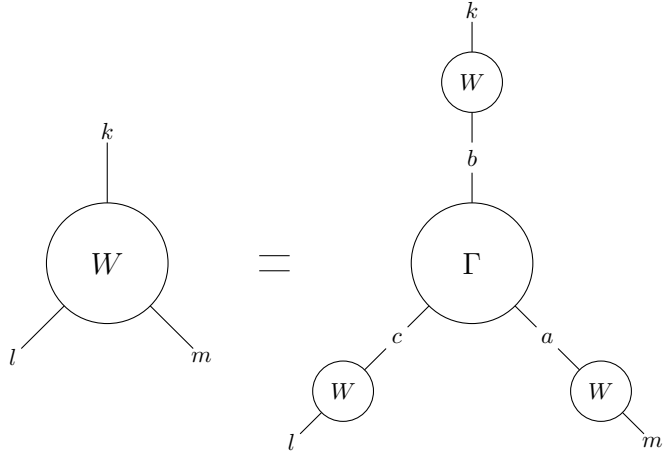
\begin{figure}[t]
\centering
    \begin{tikzpicture}[]

    \begin{scope}[scale=0.8,transform shape]

    \draw (0,0) circle (10mm);

    \draw (0,0) node {\Large $W$};

    \draw (0,1) -- (0,2);
    \draw (-0.707,-0.707) -- (-1.414,-1.414);
    \draw (0.707,-0.707) -- (1.414,-1.414);

    \draw (0,2.2) node {$k$};
    \draw (-1.55,-1.55) node {$l$};
    \draw (1.55,-1.55) node {$m$};

    \draw (2.5,0.125) -- (3,0.125);
    \draw (2.5,-0.125) -- (3,-0.125);

    \end{scope}

    \begin{scope}[scale=0.8,transform shape]

        \draw (6,0) circle (10mm);
    
        \draw (6,0) node {\Large $\Gamma$};
    
        \draw (6,1) -- (6,1.5);
        \draw (6-0.707,-0.707) -- (6-1.06,-1.06);
        \draw (6.707,-0.707) -- (7.06,-1.06);
    
        \draw (6-1.414-0.707,-1.414-0.707) circle (5mm);
        \draw (6+1.414+0.707,-1.414-0.707) circle (5mm);
        \draw (6,3) circle (5mm);
    
        \draw (6-1.414-0.707,-1.414-0.707) node {$W$};
        \draw (6+1.414+0.707,-1.414-0.707) node {$W$};
        \draw (6,3) circle node {$W$};
    
        \draw (6-1.7677-0.707,-1.7677-0.707) -- (6-2.121-0.707,-2.121-0.707);
        \draw (6+1.7677+0.707,-1.7677-0.707) -- (6+2.121+0.707,-2.121-0.707);
        \draw (6,3.5) -- (6, 4);
    
        \draw (6,4.2) node {$k$};
        \draw (6-2.26-0.707,-2.26-0.707) node {$l$};
        \draw (6+2.26+0.707,-2.26-0.707) node {$m$};
    
        \draw (6,1.75) node {$b$};
        \draw (6,2) -- (6, 2.5);
        
        \draw (6-1.06-0.177,-1.06-0.177) node {$c$};
        \draw (6-1.06-0.177-0.177,-1.414) -- (6-1.06-0.177-0.177-0.353,-1.414-0.353);
    
        \draw (6+1.06+0.177,-1.06-0.177) node {$a$};
        \draw (6+1.06+0.177+0.177,-1.414) -- (6+1.06+0.177+0.177+0.353,-1.414-0.353);
    \end{scope}
\end{tikzpicture}
\caption{A tree diagram showing the relationship between $W_{3}$ and $\Gamma_{3}$. For large enough $N$, $G_3 = S^{(3)}$.}
\label{tikz: W3}
\end{figure}

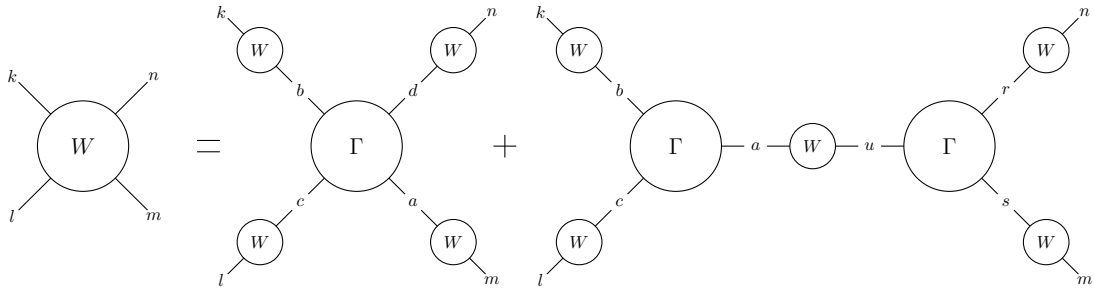
\begin{figure}[t]
\centering
    \begin{tikzpicture}[]

    \begin{scope}[scale=0.6,transform shape]

    \draw (0,0) circle (10mm);

    \draw (0,0) node {\Large $W$};

    \draw (-0.707,0.707) -- (-1.414,1.414);
    \draw (0.707,0.707) -- (1.414,1.414);
    \draw (-0.707,-0.707) -- (-1.414,-1.414);
    \draw (0.707,-0.707) -- (1.414,-1.414);

    \draw (-1.55,1.55) node {$k$};
    \draw (1.55,1.55) node {$n$};
    \draw (-1.55,-1.55) node {$l$};
    \draw (1.55,-1.55) node {$m$};

    \draw (2.5,0.125) -- (3,0.125);
    \draw (2.5,-0.125) -- (3,-0.125);

    \end{scope}

    \begin{scope}[scale=0.6,transform shape]

        \draw (6,0) circle (10mm);
    
        \draw (6,0) node {\Large $\Gamma$};
    
        \draw (6-0.707,0.707) -- (6-1.06,1.06);
        \draw (6.707,0.707) -- (7.06,1.06);
        \draw (6-0.707,-0.707) -- (6-1.06,-1.06);
        \draw (6.707,-0.707) -- (7.06,-1.06);
    
        \draw (6-1.414-0.707,-1.414-0.707) circle (5mm);
        \draw (6+1.414+0.707,-1.414-0.707) circle (5mm);
        \draw (6-1.414-0.707,1.414+0.707) circle (5mm);
        \draw (6+1.414+0.707,1.414+0.707) circle (5mm);
    
        \draw (6-1.414-0.707,-1.414-0.707) node {$W$};
        \draw (6+1.414+0.707,-1.414-0.707) node {$W$};
        \draw (6-1.414-0.707,1.414+0.707) node {$W$};
        \draw (6+1.414+0.707,1.414+0.707) node {$W$};
    
        \draw (6-1.7677-0.707,-1.7677-0.707) -- (6-2.121-0.707,-2.121-0.707);
        \draw (6+1.7677+0.707,-1.7677-0.707) -- (6+2.121+0.707,-2.121-0.707);
        \draw (6-1.7677-0.707,1.7677+0.707) -- (6-2.121-0.707,2.121+0.707);
        \draw (6+1.7677+0.707,1.7677+0.707) -- (6+2.121+0.707,2.121+0.707);
    
        \draw (6-2.26-0.707,2.26+0.707) node {$k$};
        \draw (6+2.26+0.707,2.26+0.707) node {$n$};
        \draw (6-2.26-0.707,-2.26-0.707) node {$l$};
        \draw (6+2.26+0.707,-2.26-0.707) node {$m$};

        
        \draw (6-1.06-0.177,-1.06-0.177) node {$c$};
        \draw (6-1.06-0.177-0.177,-1.414) -- (6-1.06-0.177-0.177-0.353,-1.414-0.353);
    
        \draw (6+1.06+0.177,-1.06-0.177) node {$a$};
        \draw (6+1.06+0.177+0.177,-1.414) -- (6+1.06+0.177+0.177+0.353,-1.414-0.353);

        \draw (6-1.06-0.177,1.06+0.177) node {$b$};
        \draw (6-1.06-0.177-0.177,1.414) -- (6-1.06-0.177-0.177-0.353,1.414+0.353);
    
        \draw (6+1.06+0.177,1.06+0.177) node {$d$};
        \draw (6+1.06+0.177+0.177,1.414) -- (6+1.06+0.177+0.177+0.353,1.414+0.353);

    \end{scope}

    \begin{scope}[scale=0.6,transform shape]

    \draw (9,0) -- (9.5,0);
        \draw (9.25, 0.25) -- (9.25, -0.25);

        \draw (13,0) circle (10mm);
    
        \draw (6+7,0) node {\Large $\Gamma$};
    
        \draw (6-0.707+7,0.707) -- (6+7-1.06,1.06);
        \draw (6-0.707+7,-0.707) -- (6+7-1.06,-1.06);
    
        \draw (7+6-1.414-0.707,-1.414-0.707) circle (5mm);
        \draw (7+6-1.414-0.707,1.414+0.707) circle (5mm);
    
        \draw (7+6-1.414-0.707,-1.414-0.707) node {$W$};
        \draw (7+6-1.414-0.707,1.414+0.707) node {$W$};
    
        \draw (7+6-1.7677-0.707,-1.7677-0.707) -- (7+6-2.121-0.707,-2.121-0.707);
        \draw (7+6-1.7677-0.707,1.7677+0.707) -- (7+6-2.121-0.707,2.121+0.707);
    
        \draw (7+6-2.26-0.707,2.26+0.707) node {$k$};
        \draw (7+6-2.26-0.707,-2.26-0.707) node {$l$};

        
        \draw (7+6-1.06-0.177,-1.06-0.177) node {$c$};
        \draw (7+6-1.06-0.177-0.177,-1.414) -- (7+6-1.06-0.177-0.177-0.353,-1.414-0.353);

        \draw (7+6-1.06-0.177,1.06+0.177) node {$b$};
        \draw (7+6-1.06-0.177-0.177,1.414) -- (7+6-1.06-0.177-0.177-0.353,1.414+0.353);

        \draw (14,0) -- (14.5,0);
        \draw (14.75, 0) node {$a$};
        \draw (15,0) -- (15.5,0);

        \draw (16,0) circle (5mm);
        \draw (16, 0) node {$W$};

        \draw (16.5,0) -- (17,0);
        \draw (17.25, 0) node {$u$};
        \draw (17.5,0) -- (18,0);

        \draw (19,0) circle (10mm);
    
        \draw (19,0) node {\Large $\Gamma$};
    
        \draw (19.707,0.707) -- (13+7.06,1.06);
        \draw (19.707,-0.707) -- (13+7.06,-1.06);
    
        \draw (13+6+1.414+0.707,-1.414-0.707) circle (5mm);
        \draw (13+6+1.414+0.707,1.414+0.707) circle (5mm);
    
        \draw (13+6+1.414+0.707,-1.414-0.707) node {$W$};
        \draw (13+6+1.414+0.707,1.414+0.707) node {$W$};
    
        \draw (13+6+1.7677+0.707,-1.7677-0.707) -- (13+6+2.121+0.707,-2.121-0.707);
        \draw (13+6+1.7677+0.707,1.7677+0.707) -- (13+6+2.121+0.707,2.121+0.707);
    
        \draw (13+6+2.26+0.707,2.26+0.707) node {$n$};
        \draw (13+6+2.26+0.707,-2.26-0.707) node {$m$};
    
        \draw (13+6+1.06+0.177,-1.06-0.177) node {$s$};
        \draw (13+6+1.06+0.177+0.177,-1.414) -- (13+6+1.06+0.177+0.177+0.353,-1.414-0.353);
    
        \draw (13+6+1.06+0.177,1.06+0.177) node {$r$};
        \draw (13+6+1.06+0.177+0.177,1.414) -- (13+6+1.06+0.177+0.177+0.353,1.414+0.353);


    
    \end{scope}
\end{tikzpicture}
\caption{A Tree diagram illustrating the relation between $W_4$ and $\Gamma_4$, where two symmetric terms of the last diagram are omitted. Note that the $W_3$ term that appears in eq. \eqref{eqn: W4} has already been solved for in terms of $\Gamma_3$. For large enough $N$, $\Gamma_4 = S^{(4)}$.}
\label{tikz: W4}
\end{figure}

Using a Legendre transform $\Gamma[s_c]$, we can find an easy  approximation for these functions. We may define the transform as:
\begin{equation}
    \label{eqn: Legendre transform}
    \Gamma[s_c] = -\sum_i s_c(i) j(i) +W[j].
\end{equation}
Introducing the Legendre transform is beneficial because the $n$th derivative of $\Gamma[s_c]$ (with respect to the $s_c$'s) is simply $S^{(n)}$ (at leading order in $N$). This fact, combined with the definition above, allows us to state the following:
\begin{equation}
    \label{eqn: W3}
    \begin{split}
        W_3(k,l,m) \equiv W_{klm} & = \frac{\partial^3 W}{\partial j(k)\partial j(l)\partial j(m)} = -\frac{\partial}{\partial j(m)} (\Gamma_{kl}^{-1}) = -\sum_{a}^M \frac{\partial s_c(a)}{\partial j(m)} \frac{\partial}{\partial s_c(a)} (\Gamma_{kl}^{-1}) \\
        & = \sum_{a,b,c}^M W_{am}W_{bk} \Gamma_{abc} W_{cl} \approx \sum_{a,b,c}^M W_{am}W_{bk} S^{(3)}(a,b,c) W_{cl}.
    \end{split}
\end{equation}
Above, $\Gamma^{-1}_{kl} = (\partial^2 \Gamma/\partial s_c(k) \partial s_c(l))^{-1} = -\partial^2W/\partial j(k) \partial j(l)$ - a fact that directly follows from eq. \eqref{eqn: Legendre transform}. Similarly, $W_{klmn}$ can be expressed as:
\begin{equation}
    \label{eqn: W4}
    \begin{split}
        W_4(k,l,m,n) \equiv W_{klmn} \approx & \sum_{a,b,c}^M \bigg[W_{amn} W_{bk} S^{(3)}(a,b,c) W_{cl}
        + W_{bkn} W_{am} S^{(3)}(a,b,c) W_{cl} \\
        & \hspace{1cm} + W_{cnl} W_{bk} S^{(3)}(a,b,c) W_{am} \bigg]\\
        + & \sum_{a,b,c,d}^M W_{dn} W_{am}W_{bk} S^{(4)}(a,b,c,d) W_{cl}.
    \end{split}
\end{equation}
A tree diagram representation of eqs. \eqref{eqn: W3} and \eqref{eqn: W4} can be found in figures \ref{tikz: W3} and \ref{tikz: W4}. In the derivations up until now, one might have argued that one could ignore the $1/N$ term, since we are considering the large $N$ limit. However, we included $W_4$ to account for its potentially large contribution given its complex structure.

With these substitutions, eqs. \eqref{eqn: master1} and \eqref{eqn: master 2} can be expressed entirely in terms of $W_i$ and $W_{ij}$ , drastically reducing the number of unknowns in our equations to $M + M(M+1)/2$. Since this is also the exact number of equations, either the system has one solution or none. We attempted to utilize \texttt{Solve} and \texttt{NSolve} on Mathematica to explicitly find a solution (or lack thereof). However, such a code was unable to finish running in a reasonable time, even when considering $M = 4$ and ignoring the $1/N$ terms. Thus, we turned to a numerical investigation of the 1 and 2-pt correlation functions. If the numerically evaluated values satisfy the closed equations, then the approximation is successful.

\section{Specifics of the Numerical Analysis}
\label{sec: numerical}

The correlation functions cannot be explicitly calculated, since $Z_0$ and its respective numerator diverge. Even if it were possible, such a calculation would be unreasonable for higher dimensional integrals (i.e. $M \gg 1$). This is especially troublesome for us, given that $M$ scales as a quadratic as the Euclidean lattice is taken to be a square.

\subsection{Introduction to Markov Chain Monte Carlo}

Markov Chain Mote Carlo (MCMC) is a numerical method that explicitly combats both of these issues \cite{budd2023monte, morningstar2007monte}. The fundamental strategy of this technique is to construct a Markov Chain, the mean of which is the desired value, for each observable. A simple uniform probability distribution will also yield accurate results, but the variance of the elements would fall too slowly, especially when working with many observables \cite{morningstar2007monte}. The MCMC method reduces variance by introducing two different ideas: (i) rather than uniformly choosing integration variable values, it chooses them based on a probability distribution - this is a technique called importance sampling - and (ii) element $N$ of the Markov Chain is chosen based on only element $N - 1$ and through importance sampling. Incorporating both of these properties reduces how fast the variance shrinks. The mathematics that validate this numerical technique, along with error analysis, can be found in \cite{morningstar2007monte}. 

The first of these features can be applied towards calculating integrals of the following form:
\begin{equation}
    \int_{-\infty}^\infty dx ~f(x) p(x),
\end{equation}
where $\int_{-\infty}^\infty p(x) = 1$. This $p(x)$ can be treated as a probability distribution. If we sampled points based off $p(x)$, then the integral simply yields to the mean of $f(x)$, denoted as $\langle f(x) \rangle$ with all the chosen points. This process can be trivially generalized to higher dimensions, so our calculation of $W_i$ would go as follows:
\begin{equation}
    W_i = \int [dx] ~x(i)\bigg[\frac{\exp{\left(-S\right)}}{Z_0}\bigg] = \int [dx] ~x(i) [\pi(x)] = \langle x(i) \rangle.
\end{equation}
Above, we used the fact that $\int [dx] e^{-S}/Z_0 = 1$, which makes $e^{-S}/Z_0$ a valid probability distribution $\pi(x)$ for our purposes.\footnote{We choose to use the same variable names as \cite{morningstar2007monte} throughout this manuscript to avoid confusion for readers who would like to expand their understanding of the numerical technique.} Note that we can avoid evaluating $Z_0$ altogether to show this. Similarly, $W_{ij} = \langle x(i) \rangle \langle x(j) \rangle - \langle x(i) x(j) \rangle$. 

A concrete methodology to effectively utilize MCMC is given by the Metropolis-Hastings algorithm. While this is certainly not the only algorithm, it allows us to construct an irreducible, aperiodic Markov Chain with positive recurrent states - conditions necessary for the success of MCMC \cite{morningstar2007monte}. This algorithm revolves around updating the values of the action's variable(s) (of a site) with the given probability distribution. If the action is multi variable - such as ours - then it is acceptable to either randomly choose a site to update or sequentially update each site. As stated prior, the updating process only takes in the current value $x(i)$ of a site to determine if the site can harbor the new value $\tilde{x}(i)$. That is, $P(\tilde{x} \leftarrow x)$ is only a function of $x, \tilde{x}$, and the probability distribution. 

The general outline of the algorithm is as follows \cite{budd2023monte,morningstar2007monte}:
\begin{enumerate}
    \item  We update all sites until their values `hover' over the mean. 
    \item We begin taking measurements using the thermalized values. At the start of this phase, we calculate the autocorrelation $\rho$.
    \item After obtaining many measurements, we take a subset of these as our Markov Chain based on our calculation for $\rho$.
\end{enumerate}
These steps provide the main steps of the algorithm, but the technical details (and how they relate to our program) are necessary to discuss.

\subsection{Applying Metropolis-Hasting}
\label{sec: MH}

Before we can work with the algorithm, we must numerically compute two quantities in our action: the discrete Laplacian and the stationary point. The perturbative expansion of the action with respect to $\sqrt{N}$ tells us that $\Phi_0 + (A^{-1})_{ii} = 0$ for fixed $i$. This generally needs to be solved numerically, as $A_{ij}$ is a function of $\Phi_0$. The 2D discrete Laplacian with periodic boundary conditions can be explicitly calculated using a five-point stencil on a grid like in figure \ref{tikz: grid}, but this may be tedious for larger grids. A much easier alternative is using the Kronecker sum of two 1 dimensional discrete Laplacians $L_1$ with periodic boundary conditions. Thus, the discrete Laplacian $K_{ij}$ can be written as:
\begin{equation}
    K_{ij} = L_{1} \oplus L_1 = L_1 \otimes \mathbb{1} + \mathbb{1} \otimes L_1.
    \label{eqn: laplacian}
\end{equation}
Above, $\otimes$ represents a Kronecker product. All matrices on the right hand side above are $\sqrt{M} \times \sqrt{M}$ dimensional. This forces the 2D discrete Laplacian to be an $M \times M$ matrix. When constructed like this, $K_{ij}$ may not look translation invariant. That is, $K_{ij} \neq K(i - j)$. However, a simple rearranging of rows and columns can achieve this. Doing so does not change its spectrum, so the dynamics governed by the system can be studied using $K_{ij}$ from eq. \eqref{eqn: laplacian}. There is also ambiguity in the sign of this matrix, based on the stencil chosen for $L_1$, but changing the sign simply adds a negligible constant in the action. The stationary point also changes sign, which does not affect any calculations.  

An issue with using this calculated $K_{ij}$ is the inclusion of periodic boundary conditions. This forces each row to sum to $0$, making the matrix singular. When solving $\Phi_0 + (A^{-1})_{ii} = 0$, we do not find valid solutions. Some of them are complex, while the rest are not local minimums of the action (when fluctuations are turned off).\footnote{This can be checked by plotting the action (with fluctuations turned off) as a function of $\Phi_0$.} The lack of a stationary/equilibrium phase suggests that there is no stationary point here. To combat this, we redefined $K_{ij}$ by adding an identity matrix to it. Any matrix of the form $c\mathbb{1}$ would suffice, so we chose $c = 1$ for simplicity. This rid the system of the singularity, allowing for a valid stationary point to exist. This extra term that we included can be thought of as a potential term. 


With this, the probability of updating a site's value can also be rigorously defined. As an example, let us fix all values of $x(i)$ except the one site $x(j)$ that we plan on updating. By the definition of a stochastic process (which includes a Markov Chain), a newly proposed value $\tilde{x}(j)$ for this site is accepted with the following probability:
\begin{equation}
    \label{eqn: prob update}
    P(\tilde{x}_j \leftarrow x(j)) = \min{\left(1, \frac{\pi(\tilde{x})}{\pi(x)}\right)} = \min{\left(1, \frac{e^{-S[\tilde{x}]}}{e^{-S[x]}}\right)} = \min{\left(1, \exp{\left(-(S[\tilde{x}] - S[x])\right)}\right)}.
\end{equation}
Computationally, this simply means uniformly generating a random number $p$ from $[0,1]$. If $p < P(\tilde{x}_j \leftarrow x(j))$, then the value at site $j$ is changed. If not, then the value remains the same. Regardless of if this change is accepted or rejected, we shall hereby refer to this process as `updating.'

The action that we will use in the program is the closed-form variation so that we can avoid the infinite sum. However, this introduces a subtle source of error that has big ramifications. The two versions of the action are analytically the same except for two terms: the constant term and the linear term (with respect to $x(i)$). It is evident from eq. \eqref{eqn: prob update} that the constant terms do not matter. One may be tempted to say the same thing about the linear term, but this is not true numerically. Choosing a stationary point that makes the linear term vanish is numerically equivalent to having an extremely small (and non-zero) term. Since the linear term is magnified by the corresponding factor of $\sqrt{N}$ attached to it, the desired integrals actually grow as $\sqrt{N}$. This can be avoided if we explicitly subtract out the linear term from the action. To potentially avoid unforeseen errors, we also subtracted out the constant terms.

While the action determines the likelihood of accepting a new value at a site, it does not play a role in proposing a new value at a site. There are multiple ways to do so, but in this version of the Metropolis-Hastings algorithm, we choose a point near the current value. That is, $\tilde{x}(j) = x(j) + \delta$, where $\delta$ is uniformly and randomly chosen from an interval $[-\Delta, \Delta]$. There is no quantitative formula for $\Delta$, but the consensus is to choose $\Delta$ such that the program accepts new values at sites 50\% -- 60\% of the time. This choice is to maximize efficiency and precision \cite{morningstar2007monte}. 

\begin{figure}[t]
    \centering
    \includegraphics[scale=0.3]{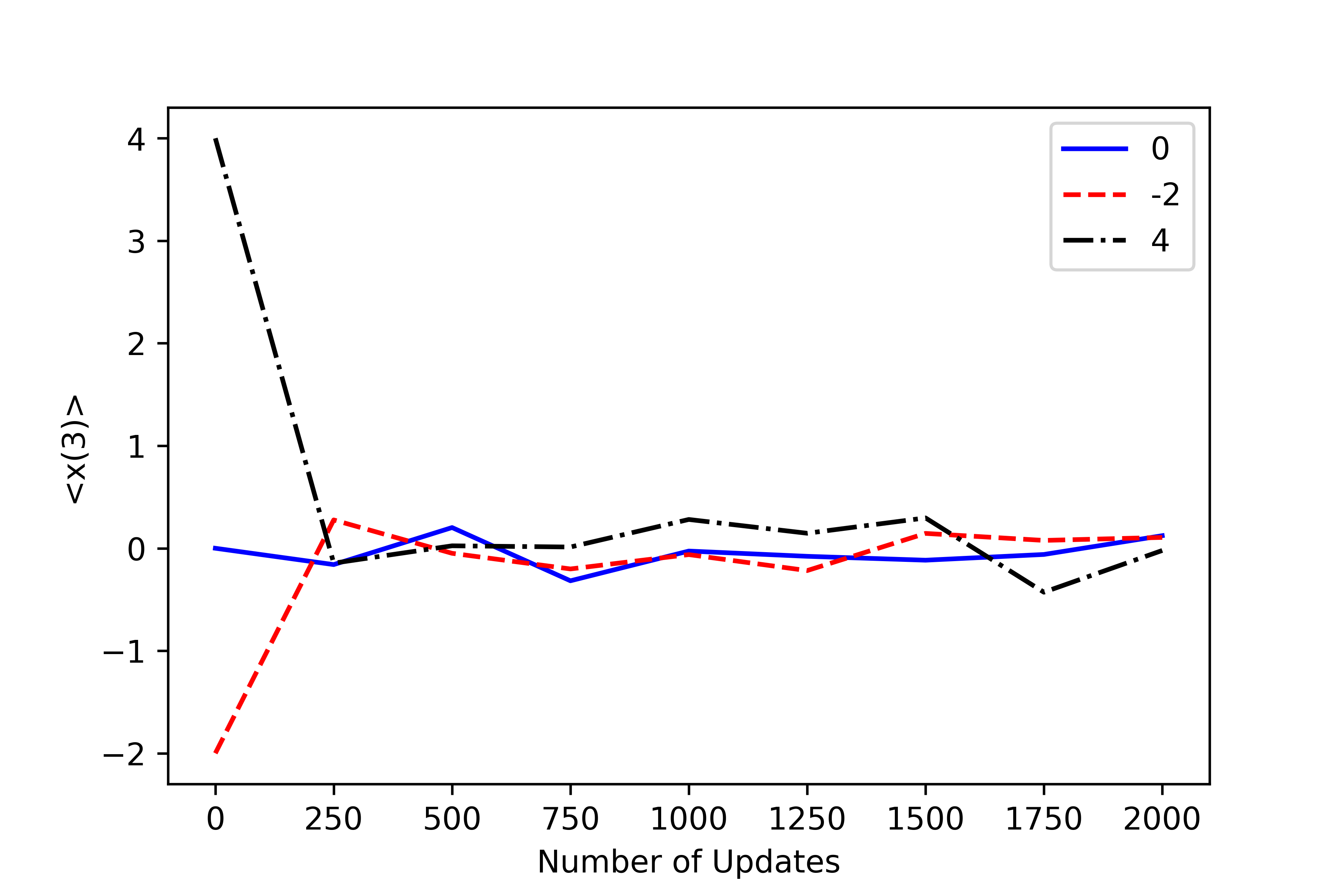}
    \caption{Plot of $\langle x_3 \rangle$ for $M = 4$ and $N = 100$. All $x(i)$ of each run started with the same values, which were $0$ (blue/solid), $-2$ (red/dashed), and $4$ (black/dot-dashed) respectively. Note that after around $\sim 300$ updates, thermalization has been achieved.}
    \label{fig:thermalization}
\end{figure}

The above method is the backbone of the main program, which begins with thermalization. Here, the site values are updated sequentially (i.e. $x(1), x(2), \cdots, x(M)$) until their means reach an indefinite phase of equilibrium. The initial values for the sites can either be all 0's (cold-start) or any value (hot-start). This choice typically does not impact the results, but the logarithm term of our action can diverge. Thus, we decided to use hot-starts with carefully chosen initial values to avoid ill-conditioned results. Reaching this state may take many sweeps of updating, where one sweep is checking every site once, but all values are not needed; only the current set of site values are needed during this phase. To determine if thermalization has been achieved, we can simply plot out the recent site value means and check if these values are near equilibrium. An example of this can  be seen in figure \ref{fig:thermalization}. At this point, we simply save the final site values to use for constructing the Markov Chain. To reduce variance, we begin the measurements phase with the average of three thermalized site values. 

The next phase is straightforward: we continuously conduct sweeps. Here, there are two simple, computational differences: (i) all previous sets of values must be stored and (ii) after each update, the observables calculated from the site values must be stored. However, only a subset of the observables' arrays will be elements of the Markov Chain. Before we discuss how to determine this subset, we must understand why all the measurements cannot be considered. By definition of a stochastic process, all measurements are dependent on the directly prior set of measurements. Since the updates are typically small, the statistical dependence that one set of measurements shares with the next few sets of measurements will bar the MCMC error from decreasing \cite{budd2023monte}. For example, consider the following integral:
\begin{equation}
    I =  \int_{\mathbb{R}} dx~x \left( \frac{e^{-x^2}}{\sqrt{\pi}} \right) = \int_{\mathbb{R}} dx~x p(x).
\end{equation}
Clearly, $I = 0$, so applying the Metropolis-Hastings algorithm should produce a Markov Chain of the form $\{x_1, -x_1,$ $x_2,$ $ -x_2,$ $ \cdots, x_n, -x_n\}$. However, updating the value of $x$ with $p(x)$ will only induce slight changes, so the transition from $x_1$ to $-x_1$ will take a couple of sweeps. These in-between values are not relevant to the integration and will only further delay prolong the error's decrease. 

The autocorrelation of the measurements can be utilized to discard these unnecessary measurements. If we were to apply Metropolis-Hasting to integrate $\int dx ~f(x) p(x)$ and obtain measurements (not Markov Chain) $Y = \{f(y_k)\}_{k = 1}^n$, then the autocorrelation between measurements $i$ and $i + h$ is defined as:
\begin{equation}
    \rho(h) = \frac{\sum_{i = 1}^{n - h}(f(y_i) - \bar{Y})(f(y_{i + h}) - \bar{Y})}{\text{Var}{(Y)}}
\end{equation}
When $\rho(h)$ is small enough, we can simply skip $h$ values in between measurements to construct our Markov Chain. There is not a consensus on this threshold, but we will use $\rho(h) < 0.1$  \cite{budd2023monte, morningstar2007monte}. When the statistical dependence of the elements is this small, the error can simply be taken to be the Var($Y_{\text{MC}}$)/$n$, where $n$ is the length of the Markov Chain $Y_{\text{MC}}$. Thus, as $n \rightarrow \infty$, the exact value of the integral is reached.  

\begin{figure}[t]
    \centering
    \includegraphics[scale=0.9]{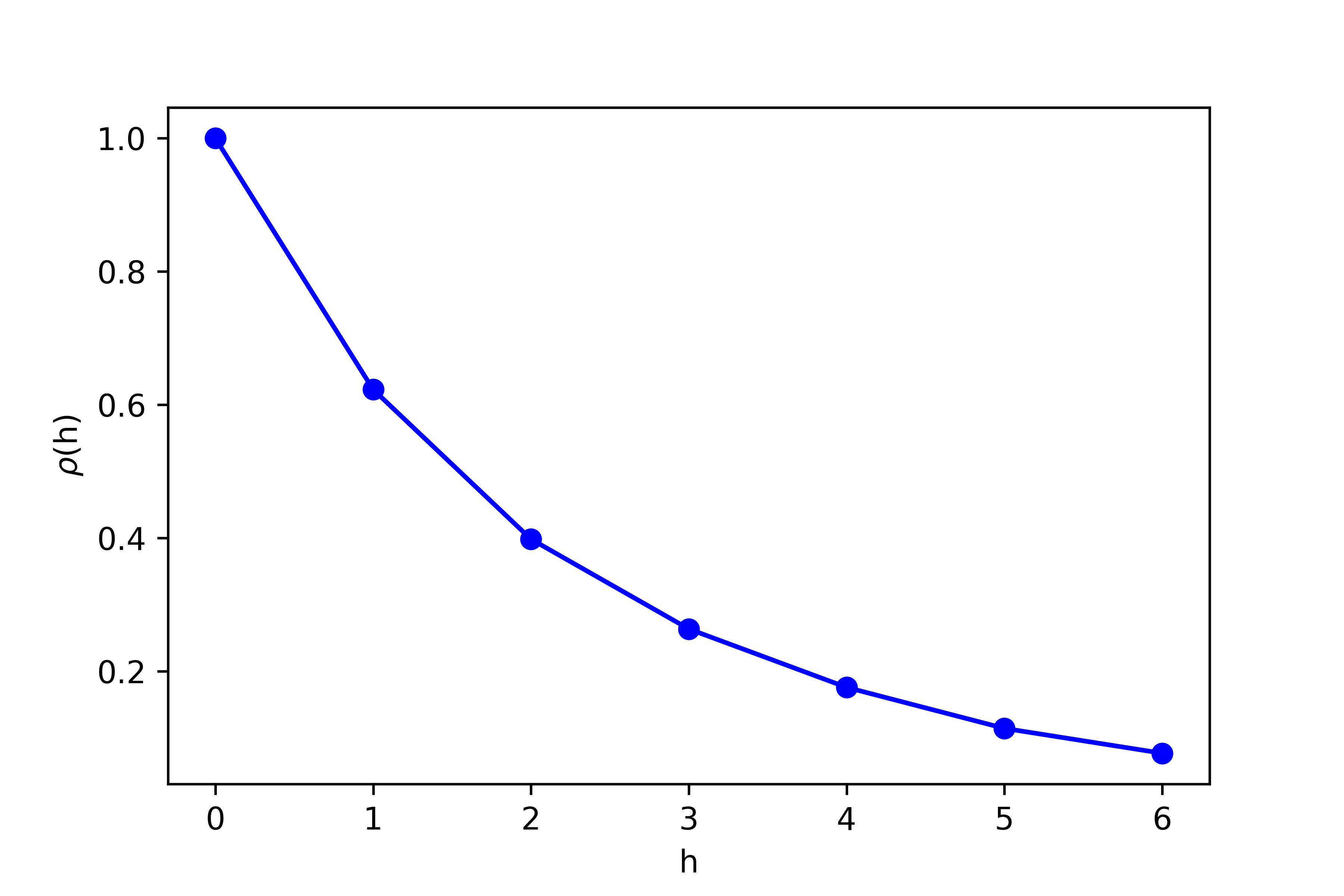}
    \caption{Plot of $\rho(h)$ with respect to $h$ for the measurements of $\langle x(1)x(4) \rangle$ for $M = 9$ and $N = 40$.}
    \label{fig:autocor}
\end{figure}

\begin{figure}[t]
    \centering
    \includegraphics[scale=0.3]{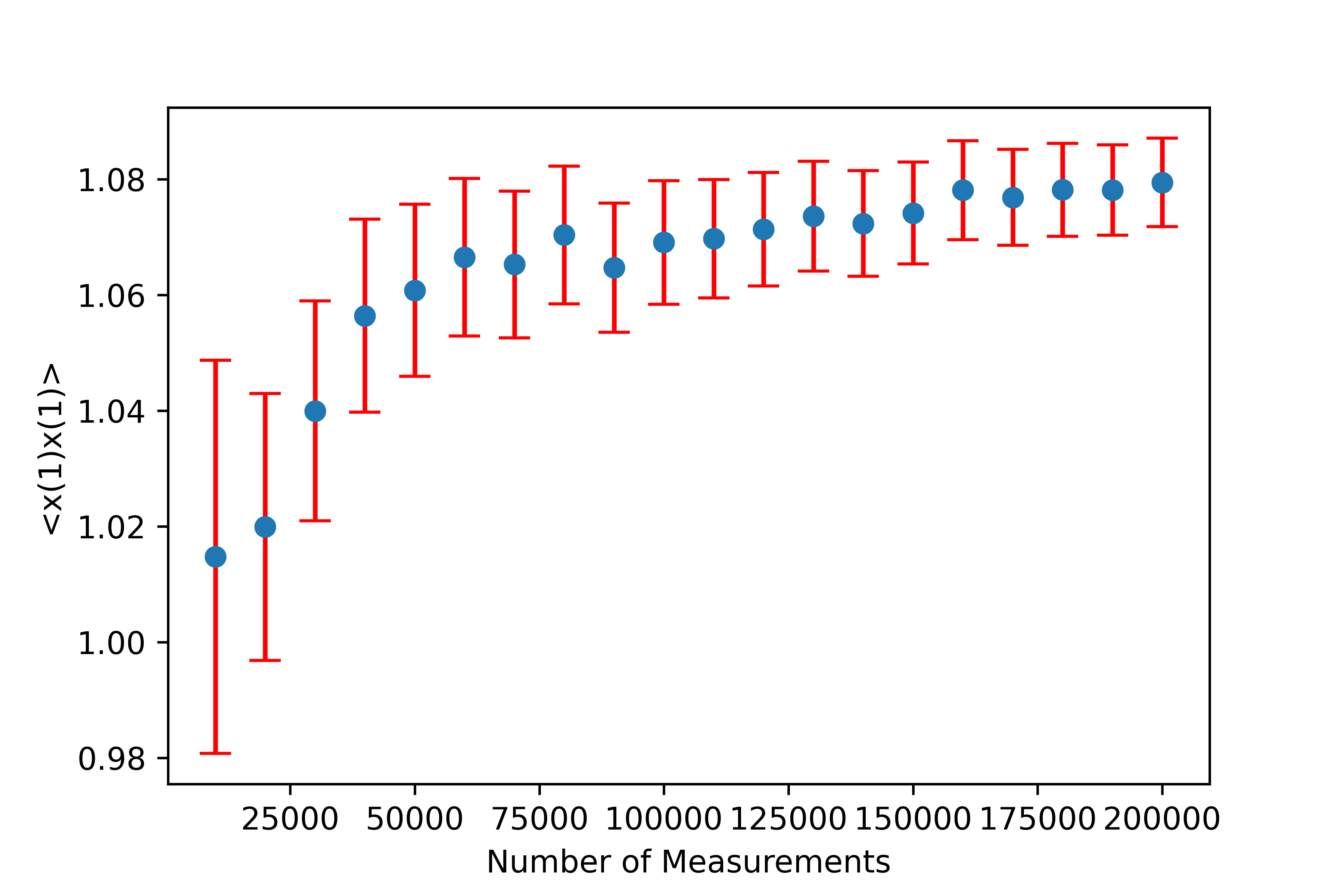}
    \caption{Plot of $\langle x(1)^2\rangle$ estimates for $N = 1000$ and $M = 9$ with error bars. The number of sweeps before adding a measurement to the Markov Chain $h_*$ was around $6$ for the $\langle x(i)^2 \rangle$ functions, including this one. Thus, the Markov Chain for this correlator was around 33,000 elements in length.}
    \label{fig:MCMC}
\end{figure}

To obtain these values in code, we update each site 10000 to obtain a large-enough array of measurements. While 10000 is an arbitrary number, a large-enough set of measurements is needed to ensure normal behavior from the variance. Then, we just keep calculating $\rho(h)$ until the first $h_*$ such that $\rho(h) < 0.1$. An example of what these typically behave like can be found in figure \ref{fig:autocor}. Each site $x(i)$ may have a slightly different $h_*$.\footnote{As far as we can tell, the length of the Markov Chain does not affect $h_*$ drastically.} We then continue the measurement phase and construct our Markov Chains with every $h_*$ measurements for each site. Figure \ref{fig:MCMC} illustrates $\langle x^2(1) \rangle$, from different numbers of measurements, along with their respective errors as an example.

It is also worth mentioning that this approximation, along with the algorithm, are easy to verify and implement for different matrices $K$ corresponding to different lattice field theories. For example, considering a discrete Laplacian that has Dirichlet boundary conditions is perfectly acceptable. The only requirement is the invertibility of the matrix, which corresponds to the existence of a valid stationary point. These points do not need to be independent of the lattice site either. Thus, this approximation and our methodology to validate it are both versatile. 

A particularly interesting problem, to which we hope to return in the future, is the application of these methods to the Homogeneous Electron Fluid. That model has a well known first order quantum phase transition between Wigner crystal and fluid phases.  Arguments have been presented that there are intermediate colloidal phases and that the transition between these and the fluid is second order.  The straightforward $1/N$ approximation cannot detect the Wigner crystal and thus misses these intermediate phases as well.  Our hope is that the resummation method described here, extended to larger lattices, can detect the full phase structure of the model.

\section{Results}
\label{sec: results}

To test our code (available on \href{https://github.com/EulersLoveChild/Markov-Chain-Monte-Carlo/tree/main}{GitHub}), we ran the code for various values of $N$ and $M = 4$ and $M = 9$, which correspond to $2 \times 2$ and $3 \times 3$ square lattices respectively (see figure \ref{tikz: grid}). For all of these $M$ and $N$ values, the value of $\Delta$, which dictates how much a site's value can change, was taken to be $2.75$. This produced a probability of acceptance of $\sim 55\%$ each time as needed. This value being the same for varying $M$ is intuitive, given that only the number of lattice sites change. On the other hand, the same fact being true for varying $N$ may seem surprising at first. Even though the effective size of the quantum fluctuation terms decreases with larger $N$, there is still an overall factor of $N$ that rectifies this downscale.

The methodology presented in section \ref{sec: numerical} is viable for higher values of $M$, but there may be some nuance to consider. Firstly, higher values of $M$ (ones larger than our examples) correspond to lattices with many more sites. Thus, $M \gg 1$ leads to the classical case. This level of discretization may make the autocorrelations fall much slower, resulting in the need for a slightly modified algorithm using microcanonical updating. Rather than updating sites with some $\delta$, the next value at the site is determined by the properties of the action in this updating technique. More information on its theory and implementation can be found in \cite{morningstar2007monte}. Other ways to reduce error can also be found in \cite{budd2023monte}. For our purposes, the number of observables grows as $M^2$, so considering $M \geq 16$ is computationally costly. This is especially true since we are running the code with $2 \times 10^5$ measurements - a value that is high enough to minimize error while being small enough to run for $M = 4$ and $M = 9$. More specifically, the error for all correlators is of order $10^{-3}$, since:
\begin{equation}
    \text{Error} = \sqrt{\frac{h_* \text{Var}(X)}{2 \times 10^5}} \approx \sqrt{\frac{(3)(0.1)}{2 \times 10^5}} = \mathcal{O}(10^{-3}).
\end{equation}
Above, we used the fact that the variance of the Markov Chain $X$ is approximately the autocorrelation threshold we set, which was $0.1$. We noticed that correlation functions of the form $\langle x^2(i) \rangle$ had larger spacing $h_*$ leading to errors high in the $10^{-3}$'s or low in the $10^{-2}$'s. 

For both cases of $M$, we choose to find correlation functions for $N = 40, 100, $ and $1000$. Larger $N$ can certainly be studied, but we found that $N = 1000$ is around the threshold in which the equations are satisfied to order $10^{-3}$. Since the error values are also at a similar level, the equations would be more prone to error and less insightful for values of $N$ larger than $1000$. On the other hand, we chose $N = 40$ as our baseline. Lower values of $N$ led to the determinant becoming $0$ or negative (from eq. \eqref{eqn: action}), when we retained a probability of acceptance of $50\% - 60\%$. We found that $N = 40$ is large enough to safely avoid this issue. 

The approximation can be verified by simply plugging in the 1 and 2-pt correlation values into the right hand sides of eqs. \eqref{eqn: master1} and \eqref{eqn: master 2}. The closer these are to $0$, the better the approximation is. The Mathematica file that we used for this purpose for $M = 9$ can be found on \href{https://github.com/EulersLoveChild/Markov-Chain-Monte-Carlo/tree/main}{GitHub}. The trend we must also look for is that these right hand side values should get smaller for larger $N$. However, we expect the error of all of our values to be of order $10^{-3}$. Thus, we must keep in mind that if the equations are well-satisfied to about order $10^{-3}$ or below, then the proposed approximation works well\footnote{This precision can be fine-tuned by increasing the number of measurements.}. When analyzing the results for these choices of $N$, we must keep in mind that we do not expect to see one drastically outperforming another. The $1/N$ term for $N = 40$ will be $\sim 25$ times larger than for $N = 1000$, and the difference between the $1/\sqrt{N}$ terms will only be off by a magnitude of $5$. This difference will be even smaller when considering the other two pairs of $N$ values. Thus, any clear decrease in the right hand sides of eqs. \eqref{eqn: master1} and \eqref{eqn: master 2} points to a successful approximation. 

A more roundabout way to check the approximation is to derive the values of the 1 and 2-pt correlators that would satisfy the leading order ($\mathcal{O}(1)$) terms of eqs. \eqref{eqn: master1} and \eqref{eqn: master 2}. Doing this reveals that $W_i$ should approach $0$ for larger $N$. This can be utilized as a simple, visual check when comparing different values of $N$. The 2-pt correlation functions are less trivial, so we will not concern ourselves with any comparisons for these functions. 


\begin{table}[t]
\centering
\resizebox{0.7\textwidth}{!}{%
\begin{tabular}{|c|c|c|c|c|c|}
\hline
 &        & $x(1)$  & $x(2)$                & $x(3)$                & $x(4)$ 
       \\ \hline
\multirow{5}{*}{$N = 40$} & $\mathbb{1}$ & $-1.35 \times 10^{-2}$ & $-1.26 \times 10^{-2}$  & $-1.18 \times 10^{-2}$  & $-1.76 \times 10^{-2}$  \\ \cline{2-6} 
 & $x(1)$ & $1.068$ & $1.49 \times 10^{-1}$ & $1.52 \times 10^{-1}$ & $2.09 \times 10^{-1}$ \\ \cline{2-6} 
 & $x(2)$ & -       & $1.073$               & $1.98 \times 10^{-1}$ & $1.48 \times 10^{-1}$ \\ \cline{2-6} 
 & $x(3)$ & -       & -                     & $1.077$               & $1.40 \times 10^{-1}$ \\ \cline{2-6} 
 & $x(4)$ & -       & -                     & -                     & $1.063$       
 \\ \hline
\multirow{5}{*}{$N = 100$} & $\mathbb{1}$ & $-9.13 \times 10^{-3}$ & $-1.11 \times 10^{-2}$ & $-2.00 \times 10^{-2}$ & $-2.11 \times 10^{-2}$ \\ \cline{2-6} 
 & $x(1)$ & $1.069$ & $1.46 \times 10^{-1}$ & $1.44 \times 10^{-1}$ & $1.97 \times 10^{-1}$ \\ \cline{2-6} 
 & $x(2)$ & -       & $1.058$               & $1.94 \times 10^{-1}$ & $1.39 \times 10^{-1}$ \\ \cline{2-6} 
 & $x(3)$ & -       & -                     & $1.063$               & $1.45 \times 10^{-1}$ \\ \cline{2-6} 
 & $x(4)$ & -       & -                     & -                     & $1.071$               \\ \hline
\multirow{5}{*}{$N = 1000$} & $\mathbb{1}$ & $-1.06 \times 10^{-2}$ & $-1.53 \times 10^{-2}$ & $-1.09 \times 10^{-2}$ & $-5.37 \times 10^{-3}$  \\ \cline{2-6} 
 & $x(1)$ & $1.060$ & $1.34 \times 10^{-1}$ & $1.39 \times 10^{-1}$ & $1.90 \times 10^{-1}$ \\ \cline{2-6} 
 & $x(2)$ & -       & $1.052$               & $1.90 \times 10^{-1}$ & $1.37 \times 10^{-1}$ \\ \cline{2-6} 
 & $x(3)$ & -       & -                     & $1.055$               & $1.41 \times 10^{-1}$ \\ \cline{2-6} 
 & $x(4)$ & -       & -                     & -                     & $1.063$                 \\ \hline
\end{tabular}%
}
\caption{Table of 1 and 2-pt correlation functions for $N = 40, 100,$ and $1000$, $M=4$, and $\Delta = 2.75$, where the element in row $x(i)$ and column $x(j)$ corresponds to $\langle x(i) x(j) \rangle$. Repeat values are not included for visibility. The error for all of these values were of order $10^{-3}$ , except the error of correlation functions $\langle x(i)^2 \rangle$. These error larger by magnitude $\sim \sqrt{2}$ due to autocorrelation spacing, so their error was close to order $10^{-2}$. 
}
\label{tbl:M = 4}
\end{table}

\subsection{\texorpdfstring{$M = 4$}{}}

The specific values of the correlation functions for $M = 4$ for $N = 40, 100$, and $1000$ can be found in table \ref{tbl:M = 4}. A straightforward calculation using the $\sqrt{N}$ term of the action revealed $\Phi_0 = -0.0709$. However, $\det{(K_{ij} + (\Phi_0 + \delta) \mathbb{1})}$ is negative, where $\delta$ here is any form of fluctuation.\footnote{Here, $K_{ij}$ is a 2D discrete Laplacian plus an identity matrix, as described in section \ref{sec: MH}.} This poses a problem as the action becomes complex. Since the region around the stationary point is the only region necessary to evaluate our desired correlators, we can instead consider the absolute value (or equivalently, changing sign, which is what we did in the code). We are simply ridding the action of a complex constant $\pm i\pi$, which is negligible.

Table \ref{tbl:results} contains what the four and ten expressions respectively evaluated to. Immediately, we can tell that the approximation works somewhat-well even at $N = 40$, but the equations are better-satisfied at higher values of $N$: there are many entries of order $10^{-2}$ for $N = 40$, but only one such entry exists for $N = 100$ and $1000$ respectively. The increased number of terms of order $10^{-4}$ and the lower coefficients for terms of order $10^{-3}$ of $N = 1000$ suggests that the approximation is better for larger $N$. Observing the 1-pt correlators in table \ref{tbl:M = 4} also indicates this; these values also get smaller for larger $N$. Both of these results strongly suggest that the approximation holds true for a $2 \times 2$ lattice from $N \sim 100$ onward.


\begin{table}[t]
\centering
\resizebox{0.7\textwidth}{!}{%
\begin{tabular}{|c|ccccc|}
\hline
\multicolumn{1}{|l|}{\multirow{2}{*}{}} &
  \multicolumn{5}{c|}{Equation Index} \\ \cline{2-6} 
\multicolumn{1}{|l|}{} &
  \multicolumn{1}{c|}{} &
  \multicolumn{1}{c|}{$1$} &
  \multicolumn{1}{c|}{$2$} &
  \multicolumn{1}{c|}{$3$} &
  $4$ \\ \hline
\multirow{5}{*}{$N = 40$} &
  \multicolumn{1}{c|}{N/A} &
  \multicolumn{1}{c|}{$4.16 \times 10^{-3}$} &
  \multicolumn{1}{c|}{$4.91 \times 10^{-3}$} &
  \multicolumn{1}{c|}{$5.96 \times 10^{-3}$} &
  $-3.58 \times 10^{-4}$ \\ \cline{2-6} 
 &
  \multicolumn{1}{c|}{$1$} &
  \multicolumn{1}{c|}{$1.25 \times 10^{-2}$} &
  \multicolumn{1}{c|}{$6.26 \times 10^{-3}$} &
  \multicolumn{1}{c|}{$2.97 \times 10^{-3}$} &
  $-5.29 \times 10^{-4}$ \\ \cline{2-6} 
 &
  \multicolumn{1}{c|}{$2$} &
  \multicolumn{1}{c|}{-} &
  \multicolumn{1}{c|}{$5.31 \times 10^{-3}$} &
  \multicolumn{1}{c|}{$1.26 \times 10^{-2}$} &
  $5.89 \times 10^{-3}$ \\ \cline{2-6} 
 &
  \multicolumn{1}{c|}{$3$} &
  \multicolumn{1}{c|}{-} &
  \multicolumn{1}{c|}{-} &
  \multicolumn{1}{c|}{$1.42 \times 10^{-3}$} &
  $1.60 \times 10^{-2}$ \\ \cline{2-6} 
 &
  \multicolumn{1}{c|}{$4$} &
  \multicolumn{1}{c|}{-} &
  \multicolumn{1}{c|}{-} &
  \multicolumn{1}{c|}{-} &
  $1.65 \times 10^{-2}$ \\ \hline
\multirow{5}{*}{$N = 100$} &
  \multicolumn{1}{c|}{N/A} &
  \multicolumn{1}{c|}{$4.63 \times 10^{-3}$} &
  \multicolumn{1}{c|}{$2.50 \times 10^{-3}$} &
  \multicolumn{1}{c|}{$-7.83 \times 10^{-3}$} &
  $-9.12 \times 10^{-3}$ \\ \cline{2-6} 
 &
  \multicolumn{1}{c|}{$1$} &
  \multicolumn{1}{c|}{$1.66 \times 10^{-3}$} &
  \multicolumn{1}{c|}{$-3.89 \times 10^{-5}$} &
  \multicolumn{1}{c|}{$2.76 \times 10^{-3}$} &
  $5.84 \times 10^{-3}$ \\ \cline{2-6} 
 &
  \multicolumn{1}{c|}{$2$} &
  \multicolumn{1}{c|}{-} &
  \multicolumn{1}{c|}{$1.17 \times 10^{-2}$} &
  \multicolumn{1}{c|}{$7.73 \times 10^{-3}$} &
  $9.39 \times 10^{-3}$ \\ \cline{2-6} 
 &
  \multicolumn{1}{c|}{$3$} &
  \multicolumn{1}{c|}{-} &
  \multicolumn{1}{c|}{-} &
  \multicolumn{1}{c|}{$7.40 \times 10^{-3}$} &
  $1.92 \times 10^{-3}$ \\ \cline{2-6} 
 &
  \multicolumn{1}{c|}{$4$} &
  \multicolumn{1}{c|}{-} &
  \multicolumn{1}{c|}{-} &
  \multicolumn{1}{c|}{-} &
  $-6.06 \times 10^{-4}$ \\ \hline
\multirow{5}{*}{$N = 1000$} &
  \multicolumn{1}{c|}{N/A} &
  \multicolumn{1}{c|}{$-5.04 \times 10^{-3}$} &
  \multicolumn{1}{c|}{$-9.80 \times 10^{-3}$} &
  \multicolumn{1}{c|}{$-4.75 \times 10^{-3}$} &
  $9.46 \times 10^{-4}$ \\ \cline{2-6} 
 &
  \multicolumn{1}{c|}{$1$} &
  \multicolumn{1}{c|}{$2.37 \times 10^{-3}$} &
  \multicolumn{1}{c|}{$5.10 \times 10^{-3}$} &
  \multicolumn{1}{c|}{$5.10 \times 10^{-4}$} &
  $5.18 \times 10^{-3}$ \\ \cline{2-6} 
 &
  \multicolumn{1}{c|}{$2$} &
  \multicolumn{1}{c|}{-} &
  \multicolumn{1}{c|}{$1.02 \times 10^{-2}$} &
  \multicolumn{1}{c|}{$4.49 \times 10^{-3}$} &
  $3.36 \times 10^{-3}$ \\ \cline{2-6} 
 &
  \multicolumn{1}{c|}{$3$} &
  \multicolumn{1}{c|}{-} &
  \multicolumn{1}{c|}{-} &
  \multicolumn{1}{c|}{$7.98 \times 10^{-3}$} &
  $-1.15 \times 10^{-3}$ \\ \cline{2-6} 
 &
  \multicolumn{1}{c|}{$4$} &
  \multicolumn{1}{c|}{-} &
  \multicolumn{1}{c|}{-} &
  \multicolumn{1}{c|}{-} &
  $-1.91 \times 10^{-4}$ \\ \hline
\end{tabular}%
}
\caption{Table of right-hand side expressions of eqs. \eqref{eqn: master1} and \eqref{eqn: master 2} with the calculated 1 and 2-pt correlation functions for $N = 40, 100,$ and $1000$. The first row of each $N$ value corresponds to the four 1st order equations, while the last four rows correspond to the 10 unique 2nd order equations. Note that the closer they are to 0, the more the aforementioned equations are true - validating the approximation. It is clear to see that the maximum value for $N$ decreases as $N$ increases, which strongly suggest that the approximation is true for large $N$.}
\label{tbl:results}
\end{table}

\subsection{\texorpdfstring{$M = 9$}{}}

For the $3 \times 3$ lattice, there are nine $1$ and forty-five $2$-pt correlation functions. As before, we considered $N = 40, 100,$ and $1000$. The stationary point for this system was found to be $\Phi_0 = 0.261$. Their values, along with what the expressions of the approximation evaluate to, can be found in table \ref{tbl: M = 9}. At an initial glance, we see that the 1-pt correlation functions decrease in magnitude and approach $0$ for larger $N$. This fact is a good indicator of our approximation's success for this lattice.

When we look at the results from eq. \eqref{eqn: master1}, we see that the $N = 40$ and $N = 1000$ cases are satisfied well. The $N = 100$ case does this to a slightly lesser extent, due to the $2$ extra entries of order $10^{-2}$ (which are right on the cusp of $10^{-3}$). Before making a conclusion, we must also consider the forty-five remaining values from eq. \eqref{eqn: master 2}. Here, we find a consistent improvement for larger $N$. Namely, the number of terms of order $10^{-2}$ and $10^{-3}$ decrease as $3 \rightarrow 3 \rightarrow 2$ and $38 \rightarrow 34 \rightarrow 32$ respectively for $N = 40, 100$, and $1000$. These missing terms appear as $4 \rightarrow 7 \rightarrow 10$ values of order $10^{-4}$ and $0 \rightarrow 1 \rightarrow 1$ values of order $10^{-5}$. 

While the first nine equation results may seem worrisome, we must keep in mind how close in proximity the values of $N$ are. This reason is why the 1 and 2-pt function values for $N = 40$ and $N = 100$ are similar, though the latter is slightly better overall. The more telling scenario is when we consider the $N = 40$ and $N = 1000$ values; the latter terms satisfy the equations much better. When assessing both sets of equations and their results, we see that there is an overall improvement for larger $N$. All of this analysis signifies that the approximation holds true on a $3 \times 3$ lattice for $N \sim 100$ and higher.


\section{Conclusion}

We've seen that the proposed level of Schwinger-Dyson (SD) truncation gives excellent results on very small square lattices. This encourages us to proceed with the main physics objective that motivated the invention of these approximation schemes. The Homogeneous Electron Fluid (HEF) is the basic model on which condensed matter physics is built. Density Functional Theory (DFT) is an attempt to write down phenomenological ansatze for the correlation functions of this field theory, in order to understand real materials. From the earliest days of study of the HEF it was realized that it had both a fluid and a crystalline phase. More recently\cite{ks} colloidal phases have been proposed, and a persuasive mean field argument for the existence of these phases has been given in two dimensions.  

In a series of papers \cite{tbbz1, tbbz2, tbbz3}, one of the authors has tried to explore these colloidal phases more systematically. In \cite{banks2020systematic}, he proposed that one could do so by searching for crystalline solutions of the truncated large $N$ SD equations.  These would provide a more systematic approach to the Wigner crystal phase than the complicated semi-classical solution of the many body Schrodinger equation and allow an extrapolation of the solution into the region where the crystal became meta-stable. As argued in \cite{tbbz1, tbbz2, tbbz3} this is the regime where one expects gel like colloids to exist. One should also be able to follow the system through its metal-insulator transition and check whether the Kivelson-Spivak prediction of crystalline impurities with negative surface tension is valid at large enough $N$. Finally, the disappearance of the crystalline solution at a finite value of the density would be a neat explanation of the mysterious phase transition unveiled in \cite{haule,japan}. 

\bibliographystyle{unsrt}
\bibliography{ref}

\newpage

\appendix

\section{\texorpdfstring{1 and 2-pt correlation function values for $M = 9$}{}}
\setcounter{table}{0}
\renewcommand{\thetable}{A\arabic{table}}

\footnotesize

\begin{longtable}[c]{|c|ccc|ccc|}
\hline
\multirow{3}{*}{Index} &
  \multicolumn{3}{c|}{Correlator Value} &
  \multicolumn{3}{c|}{Equation Results} \\ \cline{2-7} 
 &
  \multicolumn{3}{c|}{$N$} &
  \multicolumn{3}{c|}{$N$} \\ \cline{2-7} 
 &
  \multicolumn{1}{c|}{$40$} &
  \multicolumn{1}{c|}{$100$} &
  $1000$ &
  \multicolumn{1}{c|}{$40$} &
  \multicolumn{1}{c|}{$100$} &
  $1000$ \\ \hline
\endfirsthead
\endhead
$1$ &
  \multicolumn{1}{c|}{$2.14 \times 10^{-2}$} &
  \multicolumn{1}{c|}{$-7.57 \times 10^{-4}$} &
  $3.97 \times 10^{-3}$ &
  \multicolumn{1}{c|}{$7.38 \times 10^{-3}$} &
  \multicolumn{1}{c|}{$-7.86 \times 10^{-3}$} &
  $1.15 \times 10^{-3}$ \\ \hline
$2$ &
  \multicolumn{1}{c|}{$9.51 \times 10^{-3}$} &
  \multicolumn{1}{c|}{$1.87 \times 10^{-2}$} &
  $-2.06 \times 10^{-3}$ &
  \multicolumn{1}{c|}{$-4.36 \times 10^{-3}$} &
  \multicolumn{1}{c|}{$9.84 \times 10^{-3}$} &
  $-4.70 \times 10^{-3}$ \\ \hline
$3$ &
  \multicolumn{1}{c|}{$1.38 \times 10^{-2}$} &
  \multicolumn{1}{c|}{$-2.79 \times 10^{-3}$} &
  $-5.82 \times 10^{-3}$ &
  \multicolumn{1}{c|}{$-2.78 \times 10^{-4}$} &
  \multicolumn{1}{c|}{$-1.04 \times 10^{-2}$} &
  $-7.75 \times 10^{-3}$ \\ \hline
$4$ &
  \multicolumn{1}{c|}{$1.58 \times 10^{-2}$} &
  \multicolumn{1}{c|}{$6.43 \times 10^{-3}$} &
  $-1.35 \times 10^{-5}$ &
  \multicolumn{1}{c|}{$2.29 \times 10^{-3}$} &
  \multicolumn{1}{c|}{$-1.36 \times 10^{-3}$} &
  $-2.00 \times 10^{-3}$ \\ \hline
$5$ &
  \multicolumn{1}{c|}{$1.29 \times 10^{-2}$} &
  \multicolumn{1}{c|}{$1.05 \times 10^{-2}$} &
  $-1.48 \times 10^{-5}$ &
  \multicolumn{1}{c|}{$-1.30 \times 10^{-3}$} &
  \multicolumn{1}{c|}{$2.47 \times 10^{-3}$} &
  $-2.41 \times 10^{-3}$ \\ \hline
$6$ &
  \multicolumn{1}{c|}{$1.56 \times 10^{-2}$} &
  \multicolumn{1}{c|}{$-2.52 \times 10^{-3}$} &
  $9.86 \times 10^{-3}$ &
  \multicolumn{1}{c|}{$1.21 \times 10^{-3}$} &
  \multicolumn{1}{c|}{$-1.02 \times 10^{-2}$} &
  $6.86 \times 10^{-3}$ \\ \hline
$7$ &
  \multicolumn{1}{c|}{$2.40 \times 10^{-2}$} &
  \multicolumn{1}{c|}{$1.53 \times 10^{-2}$} &
  $5.34 \times 10^{-3}$ &
  \multicolumn{1}{c|}{$9.55 \times 10^{-3}$} &
  \multicolumn{1}{c|}{$6.67 \times 10^{-3}$} &
  $2.77 \times 10^{-3}$ \\ \hline
$8$ &
  \multicolumn{1}{c|}{$8.87 \times 10^{-3}$} &
  \multicolumn{1}{c|}{$-2.90 \times 10^{-3}$} &
  $-1.45 \times 10^{-3}$ &
  \multicolumn{1}{c|}{$-4.94 \times 10^{-3}$} &
  \multicolumn{1}{c|}{$-9.66 \times 10^{-3}$} &
  $-3.80 \times 10^{-3}$ \\ \hline
$9$ &
  \multicolumn{1}{c|}{$7.09 \times 10^{-3}$} &
  \multicolumn{1}{c|}{$3.51 \times 10^{-3}$} &
  $3.69 \times 10^{-3}$ &
  \multicolumn{1}{c|}{$-6.53 \times 10^{-3}$} &
  \multicolumn{1}{c|}{$-4.73 \times 10^{-3}$} &
  $1.23 \times 10^{-3}$ \\ \hline
$11$ &
  \multicolumn{1}{c|}{\textbf{$1.085$}} &
  \multicolumn{1}{c|}{\textbf{$1.082$}} &
  \textbf{$1.092$} &
  \multicolumn{1}{c|}{\textbf{$-1.18 \times 10^{-3}$}} &
  \multicolumn{1}{c|}{\textbf{$1.81 \times 10^{-5}$}} &
  \textbf{$-1.04 \times 10^{-2}$} \\ \hline
$21$ &
  \multicolumn{1}{c|}{$1.34 \times 10^{-2}$} &
  \multicolumn{1}{c|}{$1.39 \times 10^{-2}$} &
  $4.74 \times 10^{-3}$ &
  \multicolumn{1}{c|}{$-1.94 \times 10^{-3}$} &
  \multicolumn{1}{c|}{$-3.28 \times 10^{-3}$} &
  $5.39 \times 10^{-3}$ \\ \hline
$22$ &
  \multicolumn{1}{c|}{$1.091$} &
  \multicolumn{1}{c|}{$1.076$} &
  $1.073$ &
  \multicolumn{1}{c|}{$-6.70 \times 10^{-3}$} &
  \multicolumn{1}{c|}{$5.43 \times 10^{-3}$} &
  $7.43 \times 10^{-3}$ \\ \hline
$31$ &
  \multicolumn{1}{c|}{$2.02 \times 10^{-2}$} &
  \multicolumn{1}{c|}{$1.31 \times 10^{-2}$} &
  $7.97 \times 10^{-3}$ &
  \multicolumn{1}{c|}{$-8.60 \times 10^{-3}$} &
  \multicolumn{1}{c|}{$-2.34 \times 10^{-3}$} &
  $2.54 \times 10^{-3}$ \\ \hline
$32$ &
  \multicolumn{1}{c|}{$1.19 \times 10^{-2}$} &
  \multicolumn{1}{c|}{$5.79 \times 10^{-3}$} &
  $1.30 \times 10^{-2}$ &
  \multicolumn{1}{c|}{$-6.45 \times 10^{-4}$} &
  \multicolumn{1}{c|}{$4.58 \times 10^{-3}$} &
  $-2.19 \times 10^{-3}$ \\ \hline
$33$ &
  \multicolumn{1}{c|}{$1.075$} &
  \multicolumn{1}{c|}{$1.075$} &
  $1.077$ &
  \multicolumn{1}{c|}{$7.67 \times 10^{-3}$} &
  \multicolumn{1}{c|}{$6.29 \times 10^{-3}$} &
  $3.92 \times 10^{-3}$ \\ \hline
$41$ &
  \multicolumn{1}{c|}{$1.34 \times 10^{-2}$} &
  \multicolumn{1}{c|}{$4.85 \times 10^{-3}$} &
  $2.94 \times 10^{-3}$ &
  \multicolumn{1}{c|}{$-1.81 \times 10^{-3}$} &
  \multicolumn{1}{c|}{$5.59 \times 10^{-3}$} &
  $7.06 \times 10^{-3}$ \\ \hline
$42$ &
  \multicolumn{1}{c|}{$5.13 \times 10^{-2}$} &
  \multicolumn{1}{c|}{$4.10 \times 10^{-2}$} &
  $4.78 \times 10^{-2}$ &
  \multicolumn{1}{c|}{$-3.48 \times 10^{-3}$} &
  \multicolumn{1}{c|}{$5.13 \times 10^{-3}$} &
  $-1.62 \times 10^{-3}$ \\ \hline
$43$ &
  \multicolumn{1}{c|}{$4.61 \times 10^{-2}$} &
  \multicolumn{1}{c|}{$3.99 \times 10^{-2}$} &
  $4.67 \times 10^{-2}$ &
  \multicolumn{1}{c|}{$1.04 \times 10^{-3}$} &
  \multicolumn{1}{c|}{$5.47 \times 10^{-3}$} &
  $-3.37 \times 10^{-4}$ \\ \hline
$44$ &
  \multicolumn{1}{c|}{\textbf{$1.068$}} &
  \multicolumn{1}{c|}{\textbf{$1.076$}} &
  \textbf{$1.072$} &
  \multicolumn{1}{c|}{\textbf{$1.47 \times 10^{-2}$}} &
  \multicolumn{1}{c|}{\textbf{$4.72 \times 10^{-3}$}} &
  \textbf{$9.08 \times 10^{-3}$} \\ \hline
$51$ &
  \multicolumn{1}{c|}{$4.46 \times 10^{-2}$} &
  \multicolumn{1}{c|}{$4.44 \times 10^{-2}$} &
  $5.19 \times 10^{-2}$ &
  \multicolumn{1}{c|}{$2.55 \times 10^{-3}$} &
  \multicolumn{1}{c|}{$1.68 \times 10^{-3}$} &
  $-5.12 \times 10^{-3}$ \\ \hline
$52$ &
  \multicolumn{1}{c|}{$1.54 \times 10^{-3}$} &
  \multicolumn{1}{c|}{$1.56 \times 10^{-2}$} &
  $4.68 \times 10^{-3}$ &
  \multicolumn{1}{c|}{$9.15 \times 10^{-3}$} &
  \multicolumn{1}{c|}{$-4.19 \times 10^{-3}$} &
  $5.79 \times 10^{-3}$ \\ \hline
$53$ &
  \multicolumn{1}{c|}{$4.31 \times 10^{-2}$} &
  \multicolumn{1}{c|}{$3.51 \times 10^{-2}$} &
  $4.00 \times 10^{-2}$ &
  \multicolumn{1}{c|}{$3.83 \times 10^{-3}$} &
  \multicolumn{1}{c|}{$1.05 \times 10^{-2}$} &
  $5.77 \times 10^{-3}$ \\ \hline
$54$ &
  \multicolumn{1}{c|}{$1.39 \times 10^{-2}$} &
  \multicolumn{1}{c|}{$3.03 \times 10^{-3}$} &
  $2.08 \times 10^{-2}$ &
  \multicolumn{1}{c|}{$-2.61 \times 10^{-3}$} &
  \multicolumn{1}{c|}{$6.32 \times 10^{-3}$} &
  $-9.41 \times 10^{-3}$ \\ \hline
$55$ &
  \multicolumn{1}{c|}{$1.088$} &
  \multicolumn{1}{c|}{$1.071$} &
  $1.076$ &
  \multicolumn{1}{c|}{$-4.21 \times 10^{-3}$} &
  \multicolumn{1}{c|}{$1.05 \times 10^{-2}$} &
  $4.82 \times 10^{-3}$ \\ \hline
$61$ &
  \multicolumn{1}{c|}{$5.11 \times 10^{-2}$} &
  \multicolumn{1}{c|}{$4.75 \times 10^{-2}$} &
  $4.57 \times 10^{-2}$ &
  \multicolumn{1}{c|}{$-3.78 \times 10^{-3}$} &
  \multicolumn{1}{c|}{$-1.03 \times 10^{-3}$} &
  $6.64 \times 10^{-4}$ \\ \hline
$62$ &
  \multicolumn{1}{c|}{$4.62 \times 10^{-2}$} &
  \multicolumn{1}{c|}{$4.21 \times 10^{-2}$} &
  $4.65 \times 10^{-2}$ &
  \multicolumn{1}{c|}{$1.24 \times 10^{-3}$} &
  \multicolumn{1}{c|}{$3.81 \times 10^{-3}$} &
  $-7.25 \times 10^{-4}$ \\ \hline
$63$ &
  \multicolumn{1}{c|}{$1.52 \times 10^{-2}$} &
  \multicolumn{1}{c|}{$5.54 \times 10^{-3}$} &
  $4.19 \times 10^{-3}$ &
  \multicolumn{1}{c|}{$-3.28 \times 10^{-3}$} &
  \multicolumn{1}{c|}{$5.12 \times 10^{-3}$} &
  $6.46 \times 10^{-3}$ \\ \hline
$64$ &
  \multicolumn{1}{c|}{$1.82 \times 10^{-2}$} &
  \multicolumn{1}{c|}{$1.16 \times 10^{-2}$} &
  $1.58 \times 10^{-2}$ &
  \multicolumn{1}{c|}{$-6.54 \times 10^{-3}$} &
  \multicolumn{1}{c|}{$-1.13 \times 10^{-3}$} &
  $4.72 \times 10^{-3}$ \\ \hline
$65$ &
  \multicolumn{1}{c|}{$1.56 \times 10^{-2}$} &
  \multicolumn{1}{c|}{$1.28 \times 10^{-2}$} &
  $9.25 \times 10^{-3}$ &
  \multicolumn{1}{c|}{$-4.49 \times 10^{-3}$} &
  \multicolumn{1}{c|}{$-1.70 \times 10^{-3}$} &
  $1.28 \times 10^{-3}$ \\ \hline
$66$ &
  \multicolumn{1}{c|}{\textbf{$1.079$}} &
  \multicolumn{1}{c|}{\textbf{$1.081$}} &
  \textbf{$1.086$} &
  \multicolumn{1}{c|}{\textbf{$5.39 \times 10^{-3}$}} &
  \multicolumn{1}{c|}{\textbf{$4.93 \times 10^{-4}$}} &
  \textbf{$-4.14 \times 10^{-3}$} \\ \hline
$71$ &
  \multicolumn{1}{c|}{$7.82 \times 10^{-3}$} &
  \multicolumn{1}{c|}{$7.67 \times 10^{-3}$} &
  $8.72 \times 10^{-3}$ &
  \multicolumn{1}{c|}{$4.05 \times 10^{-3}$} &
  \multicolumn{1}{c|}{$3.24 \times 10^{-3}$} &
  $1.98 \times 10^{-3}$ \\ \hline
$72$ &
  \multicolumn{1}{c|}{$4.40 \times 10^{-2}$} &
  \multicolumn{1}{c|}{$4.03 \times 10^{-2}$} &
  $4.59 \times 10^{-2}$ &
  \multicolumn{1}{c|}{$2.97 \times 10^{-3}$} &
  \multicolumn{1}{c|}{$5.77 \times 10^{-3}$} &
  $-1.11 \times 10^{-5}$ \\ \hline
$73$ &
  \multicolumn{1}{c|}{$4.24 \times 10^{-2}$} &
  \multicolumn{1}{c|}{$5.33 \times 10^{-2}$} &
  $4.92 \times 10^{-2}$ &
  \multicolumn{1}{c|}{$4.39 \times 10^{-3}$} &
  \multicolumn{1}{c|}{$-7.40 \times 10^{-3}$} &
  $-3.23 \times 10^{-3}$ \\ \hline
$74$ &
  \multicolumn{1}{c|}{$6.30 \times 10^{-3}$} &
  \multicolumn{1}{c|}{$1.81 \times 10^{-3}$} &
  $1.08 \times 10^{-2}$ &
  \multicolumn{1}{c|}{$5.31 \times 10^{-3}$} &
  \multicolumn{1}{c|}{$7.93 \times 10^{-3}$} &
  $7.41 \times 10^{-4}$ \\ \hline
$75$ &
  \multicolumn{1}{c|}{$4.67 \times 10^{-2}$} &
  \multicolumn{1}{c|}{$4.23 \times 10^{-2}$} &
  $4.09 \times 10^{-2}$ &
  \multicolumn{1}{c|}{$3.93 \times 10^{-4}$} &
  \multicolumn{1}{c|}{$3.61 \times 10^{-3}$} &
  $4.46 \times 10^{-3}$ \\ \hline
$76$ &
  \multicolumn{1}{c|}{$5.37 \times 10^{-2}$} &
  \multicolumn{1}{c|}{$3.69 \times 10^{-2}$} &
  $4.99 \times 10^{-2}$ &
  \multicolumn{1}{c|}{$-5.69 \times 10^{-3}$} &
  \multicolumn{1}{c|}{$8.72 \times 10^{-3}$} &
  $-3.33 \times 10^{-3}$ \\ \hline
$77$ &
  \multicolumn{1}{c|}{$1.094$} &
  \multicolumn{1}{c|}{$1.080$} &
  $1.080$ &
  \multicolumn{1}{c|}{$-9.50 \times 10^{-3}$} &
  \multicolumn{1}{c|}{$1.48 \times 10^{-3}$} &
  $1.48 \times 10^{-3}$ \\ \hline
$81$ &
  \multicolumn{1}{c|}{$3.75 \times 10^{-2}$} &
  \multicolumn{1}{c|}{$4.49 \times 10^{-2}$} &
  $4.38 \times 10^{-2}$ &
  \multicolumn{1}{c|}{$9.50 \times 10^{-3}$} &
  \multicolumn{1}{c|}{$1.42 \times 10^{-3}$} &
  $2.37 \times 10^{-3}$ \\ \hline
$82$ &
  \multicolumn{1}{c|}{$1.53 \times 10^{-2}$} &
  \multicolumn{1}{c|}{$1.07 \times 10^{-2}$} &
  $8.62 \times 10^{-3}$ &
  \multicolumn{1}{c|}{$-3.55 \times 10^{-3}$} &
  \multicolumn{1}{c|}{$-1.65 \times 10^{-4}$} &
  $2.07 \times 10^{-3}$ \\ \hline
$83$ &
  \multicolumn{1}{c|}{$4.91 \times 10^{-2}$} &
  \multicolumn{1}{c|}{$4.57 \times 10^{-2}$} &
  $5.03 \times 10^{-2}$ &
  \multicolumn{1}{c|}{$-1.75 \times 10^{-3}$} &
  \multicolumn{1}{c|}{$1.21 \times 10^{-4}$} &
  $-4.14 \times 10^{-3}$ \\ \hline
$84$ &
  \multicolumn{1}{c|}{$4.14 \times 10^{-2}$} &
  \multicolumn{1}{c|}{$5.39 \times 10^{-2}$} &
  $4.63 \times 10^{-2}$ &
  \multicolumn{1}{c|}{$5.05 \times 10^{-3}$} &
  \multicolumn{1}{c|}{$-7.65 \times 10^{-3}$} &
  $-3.55 \times 10^{-4}$ \\ \hline
$85$ &
  \multicolumn{1}{c|}{$8.15 \times 10^{-3}$} &
  \multicolumn{1}{c|}{$1.22 \times 10^{-2}$} &
  $7.98 \times 10^{-3}$ &
  \multicolumn{1}{c|}{$2.91 \times 10^{-3}$} &
  \multicolumn{1}{c|}{$-1.77 \times 10^{-3}$} &
  $2.99 \times 10^{-3}$ \\ \hline
$86$ &
  \multicolumn{1}{c|}{$4.58 \times 10^{-2}$} &
  \multicolumn{1}{c|}{$4.45 \times 10^{-2}$} &
  $4.33 \times 10^{-2}$ &
  \multicolumn{1}{c|}{$1.61 \times 10^{-3}$} &
  \multicolumn{1}{c|}{$1.69 \times 10^{-3}$} &
  $2.94 \times 10^{-3}$ \\ \hline
$87$ &
  \multicolumn{1}{c|}{$9.96 \times 10^{-3}$} &
  \multicolumn{1}{c|}{$8.50 \times 10^{-3}$} &
  $1.53 \times 10^{-2}$ &
  \multicolumn{1}{c|}{$1.14 \times 10^{-3}$} &
  \multicolumn{1}{c|}{$1.72 \times 10^{-3}$} &
  $-3.94 \times 10^{-3}$ \\ \hline
$88$ &
  \multicolumn{1}{c|}{\textbf{$1.070$}} &
  \multicolumn{1}{c|}{\textbf{$1.082$}} &
  \textbf{$1.110$} &
  \multicolumn{1}{c|}{\textbf{$1.19 \times 10^{-2}$}} &
  \multicolumn{1}{c|}{\textbf{$-1.65 \times 10^{-4}$}} &
  \textbf{$-2.65 \times 10^{-2}$} \\ \hline
$91$ &
  \multicolumn{1}{c|}{$4.21 \times 10^{-2}$} &
  \multicolumn{1}{c|}{$3.77 \times 10^{-2}$} &
  $4.39 \times 10^{-2}$ &
  \multicolumn{1}{c|}{$4.82 \times 10^{-3}$} &
  \multicolumn{1}{c|}{$8.08 \times 10^{-3}$} &
  $2.39 \times 10^{-3}$ \\ \hline
$92$ &
  \multicolumn{1}{c|}{$4.66 \times 10^{-2}$} &
  \multicolumn{1}{c|}{$5.14 \times 10^{-2}$} &
  $4.96 \times 10^{-2}$ &
  \multicolumn{1}{c|}{$5.34 \times 10^{-4}$} &
  \multicolumn{1}{c|}{$-4.55 \times 10^{-3}$} &
  $-3.70 \times 10^{-3}$ \\ \hline
$93$ &
  \multicolumn{1}{c|}{$1.57 \times 10^{-2}$} &
  \multicolumn{1}{c|}{$5.96 \times 10^{-3}$} &
  $1.14 \times 10^{-2}$ &
  \multicolumn{1}{c|}{$-3.96 \times 10^{-3}$} &
  \multicolumn{1}{c|}{$3.95 \times 10^{-3}$} &
  $-7.09 \times 10^{-4}$ \\ \hline
$94$ &
  \multicolumn{1}{c|}{$4.52 \times 10^{-2}$} &
  \multicolumn{1}{c|}{$3.51 \times 10^{-2}$} &
  $4.68 \times 10^{-2}$ &
  \multicolumn{1}{c|}{$1.47 \times 10^{-3}$} &
  \multicolumn{1}{c|}{$9.74 \times 10^{-3}$} &
  $-6.39 \times 10^{-4}$ \\ \hline
$95$ &
  \multicolumn{1}{c|}{$4.38 \times 10^{-2}$} &
  \multicolumn{1}{c|}{$5.76 \times 10^{-2}$} &
  $4.54 \times 10^{-2}$ &
  \multicolumn{1}{c|}{$2.84 \times 10^{-3}$} &
  \multicolumn{1}{c|}{$-1.08 \times 10^{-2}$} &
  $9.10 \times 10^{-4}$ \\ \hline
$96$ &
  \multicolumn{1}{c|}{$1.22 \times 10^{-2}$} &
  \multicolumn{1}{c|}{$1.07 \times 10^{-2}$} &
  $1.13 \times 10^{-2}$ &
  \multicolumn{1}{c|}{$-2.96 \times 10^{-4}$} &
  \multicolumn{1}{c|}{$2.33 \times 10^{-4}$} &
  $-1.43 \times 10^{-4}$ \\ \hline
$97$ &
  \multicolumn{1}{c|}{$1.35 \times 10^{-2}$} &
  \multicolumn{1}{c|}{$6.35 \times 10^{-3}$} &
  $5.09 \times 10^{-3}$ &
  \multicolumn{1}{c|}{$-2.34 \times 10^{-3}$} &
  \multicolumn{1}{c|}{$3.56 \times 10^{-3}$} &
  $5.26 \times 10^{-3}$ \\ \hline
$98$ &
  \multicolumn{1}{c|}{$8.70 \times 10^{-3}$} &
  \multicolumn{1}{c|}{$1.24 \times 10^{-2}$} &
  $1.16 \times 10^{-2}$ &
  \multicolumn{1}{c|}{$1.78 \times 10^{-3}$} &
  \multicolumn{1}{c|}{$-9.93 \times 10^{-4}$} &
  $-7.06 \times 10^{-4}$ \\ \hline
$99$ &
  \multicolumn{1}{c|}{$1.097$} &
  \multicolumn{1}{c|}{$1.081$} &
  $1.072$ &
  \multicolumn{1}{c|}{$-1.28 \times 10^{-2}$} &
  \multicolumn{1}{c|}{$-6.97 \times 10^{-4}$} &
  $8.57 \times 10^{-3}$ \\ \hline
\caption{Table of numerically evaluated 1 and 2-pt correlation functions for $M = 9$ and $N = 40, 100,$ and $1000$, along with the right hand side expressions of eqs. \eqref{eqn: master1} and \eqref{eqn: master 2} with these values plugged in. Index $k$ corresponds to $\langle x(k) \rangle$ in the first three columns and eq. \eqref{eqn: master1} with $i = k$ in the last three columns. Similarly, index $kl$ corresponds to $\langle x(k)x(l) \rangle$ in the first three columns and eq. \eqref{eqn: master1} with $i = k$ and $t = l$ in the last three columns. It is easy to that the 1-pt functions and the equation results, on average, grow closer to $0$ for larger $N$.}
\label{tbl: M = 9}
\end{longtable}

\normalsize

\end{document}